# 3D simulation of complex shading affecting PV systems taking benefit from the power of graphics cards developed for the video game industry

Jesús Robledo[1,*], Jonathan Leloux[1,2], Eduardo Lorenzo[1,2]
[1] WebPV, Madrid, Spain.
[2] Instituto de Energía Solar – Universidad Politécnica de Madrid (IES-UPM), Madrid, Spain.
*Corresponding author: jesus.robledo@webpv.net

ABSTRACT: Shading reduces the power output of a photovoltaic (PV) system. The design engineering of PV systems requires modeling and evaluating shading losses. Some PV systems are affected by complex shading scenes whose resulting PV energy losses are very difficult to evaluate with current modeling tools. Several specialized PV design and simulation software include the possibility to evaluate shading losses. They generally possess a Graphical User Interface (GUI) through which the user can draw a 3D shading scene, and then evaluate its corresponding PV energy losses. The complexity of the objects that these tools can handle is relatively limited. We have created a software solution, 3DPV, which allows evaluating the energy losses induced by complex 3D scenes on PV generators. The 3D objects can be imported from specialized 3D modeling software or from a 3D object library. The shadows cast by this 3D scene on the PV generator are then directly evaluated from the Graphics Processing Unit (GPU). Thanks to the recent development of GPUs for the video game industry, the shadows can be evaluated with a very high spatial resolution that reaches well beyond the PV cell level, in very short calculation times. A PV simulation model then translates the geometrical shading into PV energy output losses. 3DPV has been implemented using WebGL, which allows it to run directly from a Web browser, without requiring any local installation from the user. This also allows taken full benefits from the information already available from Internet, such as the 3D object libraries. This contribution describes, step by step, the method that allows *3DPV* to evaluate the PV energy losses caused by complex shading.  We then illustrate the results of this methodology to several application cases that are encountered in the world of PV systems design.

## 1 INTRODUCTION

Shading reduces the power output of a photovoltaic (PV) system. The design engineering of PV systems requires modeling and evaluating shading losses. Some PV systems are affected by complex shading scenes whose resulting PV energy losses are very difficult to evaluate with current modeling tools.

Building Integrated photovoltaics (BIPV) are sometimes exposed to shading caused by surrounding buildings, trees and lighting poles [1].

PV plants are sometimes equipped with solar trackers of complex geometry and tracking strategies, and distributed on uneven ground, crossed by electrical pylons, and bordered by mountains [2].

Several specialized PV design and simulation software include the possibility to evaluate shading losses [3,4]. They generally possess a Graphical User Interface (GUI) through which the user can draw a 3D shading scene, and then evaluate its corresponding PV energy losses. The complexity of the objects that these tools can handle is relatively limited, mainly because they evaluate the shadows cast on the PV generators using one of these two methods, which are directly executed from the Central Processing Unit (CPU):

- *Projection algorithms* [5]. They are usually restricted to 3D scenes that are made of a low number of triangles.
- *Ray tracing algorithms* [6]. They can handle very complex shading scenarios. Their accuracy and the system resource that they require depend on the number of rays that need to be traced, which can grow very fast with the complexity of the scene.

Additionally, the GUI of these tools is not as user-friendly as the one of specialized 3D modeling software.

Several specialized 3D modeling software allow modelling complex 3D scenes and rendering shading situations [7,8]. But they do not allow translating the shading that is visualized into PV energy losses.

It could be advantageous for the solar engineer to be able to model a complex 3D scene in a specialized 3D modeling software, and then import this 3D scene into a specialized PV software that is able to evaluate the PV energy losses caused by the shading corresponding to this 3D scene.

We have therefore developed a software solution, *3DPV* [9], which allows evaluating the energy losses induced by complex 3D scenes on PV generators. The 3D objects can be imported from specialized 3D modeling software or from a 3D object library. The shadows cast by this 3D scene on the PV generator are then directly evaluated by the Graphics Processing Unit (GPU). Thanks to the recent development of GPUs for the video game industry, the shadows can be evaluated with a very high spatial resolution that reaches well beyond the PV cell level, in very short calculation times. A PV simulation model then translates the geometrical shading into PV energy output losses.

*3DPV* has been implemented using *WebGL* [10], which allows it to run directly from a Web browser, without requiring any local installation from the user. This also allows taken full benefits from the information already available from Internet, such as the 3D object libraries.

This contribution describes, step by step, the method that allows *3DPV* to evaluate the PV energy losses caused by complex shading.

We then illustrate the results of this methodology to several application cases that are encountered in the world of PV systems design.

## 2 METHODOLOGY

3DPV evaluates the PV energy output losses caused by shading through a procedure that has been patented [11] and that follows four main steps:
1) Import the scene from specialized 3D modeling software or 3D object library.
2) Define the PV system with its geometry and its electrical characteristics.
3) Evaluate the shadows cast on the PV system.
4) Translate the shading affecting the PV system into energy output losses.

STEP 1: Import the 3D scene from software or library
There are several ways to create the 3D scene:
- Model it using specialized 3D modeling software, such as SketchUp [7] or AutoCAD [8] and then import it. This allows drawing the 3D objects using a user-friendly GUI. This also allows importing other kinds of objects such as topographical maps or near-horizon profiles.
- Import the 3D scene directly from a 3D object library, such as the Google Library or from *Google Earth* [12]. This allows taking benefit from the vast amount of 3D objects and geo-information already available from Internet.
- The GUI of *3DPV* also allows the user to directly define simple objects such as planes or blocks.
- All the 3D objects can also be manually moved, scaled, rotated or hidden by the user.

The user can visualize the shading scene directly into the GUI, and from any viewing direction. The 3D rendering of the objects and of the shadows is carried out using the shadow mapping technique [13].

Figure 1 shows a simple 3D scene. The wall has been directly modeled from the GUI of *3DPV*. The bike (so typical of Amsterdam) is obtained from the *Google SketchUp 3D Warehouse* online 3D object library [14]. The PV generator is defined in STEP 2.

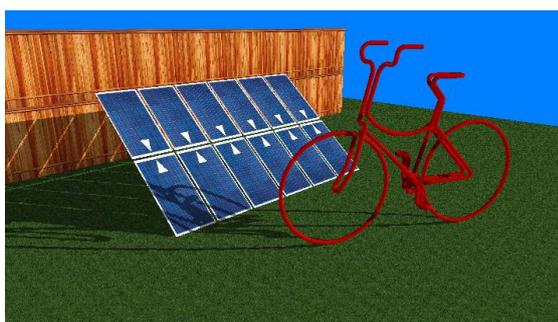

**Figure 1:** Simple 3D scene including a wall, a bike, and PV generator.

STEP 2: Define the PV system
The PV system is defined by its geometry and by its electrical characteristics. The geometry of a PV generator is defined by its dimensions, position and orientation. Its electrical characteristics are defined by the number of PV modules, their position and orientation within the PV generator, and their connection in series or parallel within each string of the PV generator. The electrical characteristics also encompass the number of cells composing each PV module, their position within the PV modules, their connection in series or parallel, and their relationship with the by-pass diodes equipping the PV modules. Figure 2 shows a PV generator composed of 12 PV modules connected in series, each of them composed of 36 cells and two bypass diodes.

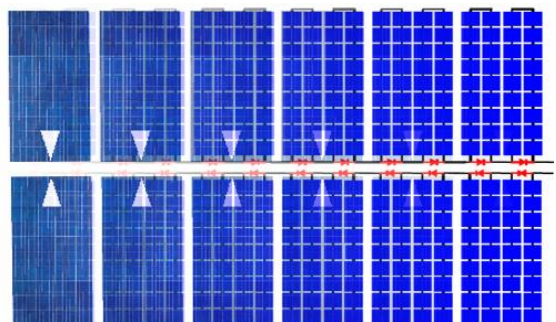

**Figure 2:** PV generator with 12 PV modules connected in series, each one with 36 cells and two bypass diodes.

STEP 3: Evaluate the shadows cast on the PV system
We need to evaluate the shadows cast by each one of the objects on each part of the PV generator, for each moment (or solar position). Our software evaluates the shading losses using the built-in z-coordinate feature from the GPU. The main graphical function of the processor is to take specifications of graphical primitives (such as lines, circles, and polygons) generated by application programs and to assign values to the pixels in the frame buffer that best represent these entities [15]. The conversion of geometric entities to pixel colors and locations in the frame buffer is known as rasterization. In early graphics systems, the frame buffer was part of the standard memory that could be directly addressed by the CPU. Today, virtually all graphics systems are characterized by special-purpose GPUs, custom-tailored to carry out specific graphics functions. The GPU can be either on the mother board of the system or on a graphics card. GPUs have evolved to where they are as complex as or even more complex than CPUs. They are characterized by both special-purpose modules geared toward graphical operations and a high degree of parallelism. GPUs are so powerful that they can often be used as mini supercomputers for general purpose computing.

One of the functionalities of the GPU during rasterization is to evaluate the depth of each pixel to be displayed for a specified field of view, defined by a source point, and direction view aperture as can be seen in figure 3. This information is stored as z-coordinate in the depth buffer.

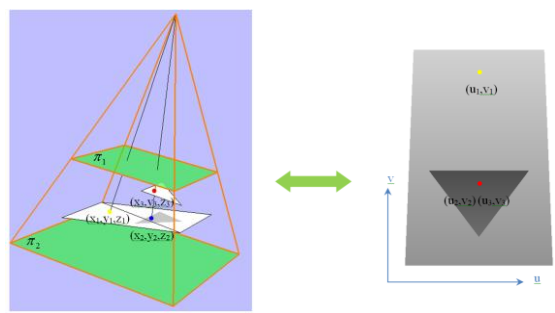

**Figure 3:** z-coordinate and depth buffer definition in the GPU. Points 2 and 3 are located in the same pixel but depth stored corresponds to point 3.

For a specific date, time and location, the solar position is defined and an orthographic view (parallel rays, in opposition to the conical view shown in Figure 3) is drawn of the 3D scene from the Sun, with a field of view limited to the PV generator of interest. While drawing the scene, the GPU stores the z-coordinate values for each pixel in the picture (shown as a grey scale in Figure 4, darker as closer to the Sun). A depth check is done for each point of interest in the PV generator. In the example of Figure 4, for simplicity, the center of each cell is considered. The points that pass the depth check (the z-coordinate corresponds to the specific pixel in the depth picture) are considered lighted (green points). The ones that fail the test (z-coordinate higher than the specific pixel's in the depth picture) are considered shaded (red points).

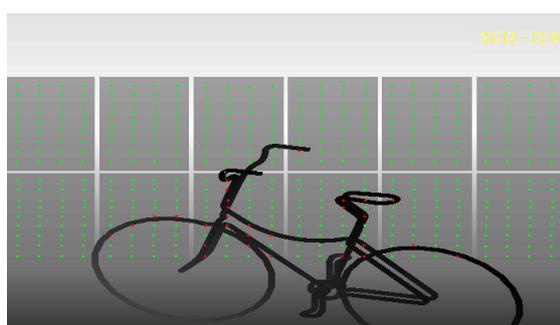

**Figure 4:** Shadow of the bike on the PV generator, visualized in field depth view from the Sun. The red points are considered as shaded by the bike because the z-coordinate of their cell is lower than the z-coordinate of the bike. The green points are considered unshaded.

We divide the PV generator into small areas, and we assign a shading status to each one of them: shaded or unshaded. The size and number of these areas can be made as small as desired. For most of the case, we have nevertheless observed that a good compromise between accuracy and computation time is reached when dividing each PV cell into 9 areas (3x3).

STEP 4: Translate the shading into PV energy losses

We estimate the solar irradiation data reaching the PV generator in absence of shading. This can be achieved, depending on the aim of the study, through:
- The input of Typical Meteorological Year (TMY) solar irradiation data.
- The input of real-time series of solar irradiance.
- The simulation of the solar irradiance with a clear-sky model [16,17].

We take away the direct component of solar irradiation from all the areas of the PV generator that were evaluated as shaded. This leads to assess the geometrical shading.

The calculation of the effective shading losses takes into account the electrical characteristics of the PV system, including its cells and diodes configuration. We use a PV diode model [18] and a PV system performance model [19] to translate the geometrical shading into a loss in PV energy output.

These shading losses evaluations can be carried out at any temporal resolution. We have come to the conclusion that for most of the applications, a temporal resolution of 1 simulation each 10 minutes leads to a good compromise between accuracy and calculation time.

The PV energy output losses evaluated for each moment are then integrated to quantify the losses corresponding to longer intervals of time, such as one day, one month, or one year.

## 3 RESULTS

We illustrate the results that are achievable with 3DPV through its application to the evaluation of different shading scenes that are representative of a diversified range of complex shading situations that PV engineers face in the real world of PV systems design and simulation. For each example, a short explanation of the functionality demonstrated is described.

### 3.1 BIPV on a house shaded by a tree and a streetlight

Figure 5 shows a PV generator installed on the roof of a house, on which are casted the shadows of a tree and a streetlight. The house, the tree and the streetlight were downloaded from the Google 3D Warehouse online object library. These 3D objects possess very complex shapes and as a result, the complete scene was composed of approximately 80,000 triangles.

The PV generator was defined as composed of 39 panels (3x13 matrix), each of them made of PV 36 cells.

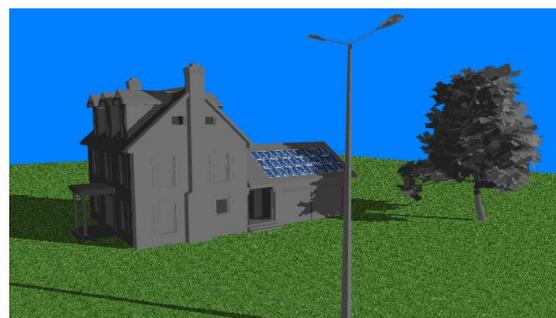

**Figure 5:** PV generator installed on the roof of a house, on which are casted the shadows of a tree and a streetlight.

Figure 6 shows a heat map representing the effective shading losses on a 2D Sun's trajectory diagram (azimuth – zenith). We appreciate the shading losses due to the house during the summer evenings, and the shading losses due to the tree during the winter mornings. The shading losses caused by the streetlight during the winter mornings are also visible.

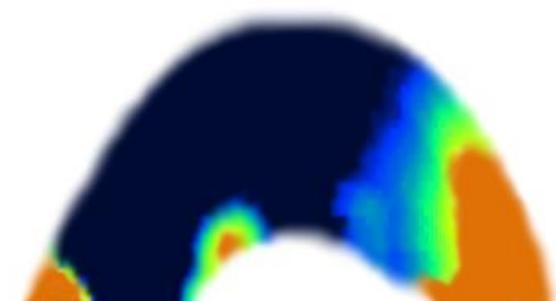

**Figure 6:** Heat map for effective shading factor between summer and winter solstice.

3.2 BIPV on the roof of the Amsterdam RAI

Figure 7 shows the RAI exhibition and convention center of Amsterdam. The 3D models were downloaded from *Google Earth*. They were positioned at their real locations and orientations. A hypothetical PV system was mounted on the roof. The shading was then directly evaluated.

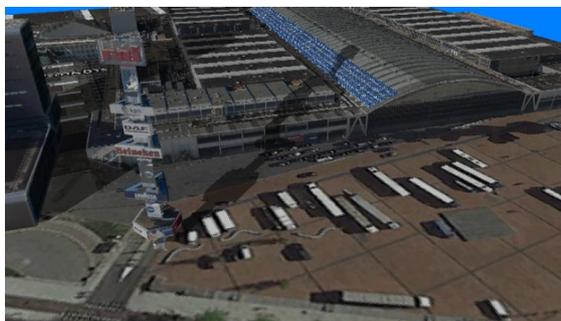

**Figure 7:** RAI exhibition and convention center of Amsterdam.

3.3 BIPV on the football stadium of Mineirao in Brazil

Figure 8 shows the real case of a project that we carried out in June 2012 to evaluate the shading on the PV generator that was due to be installed on the Mineirao stadium in Brazil, for the football FIFA World Cup 2014.

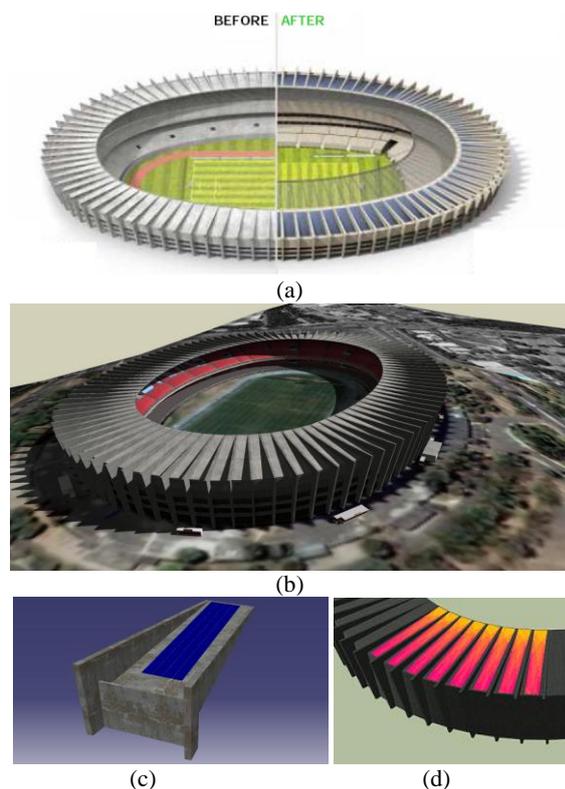

**Figure 8:** *(a)* The Mineirao stadium before and after the installation of the PV generator. *(b)* 3D model of the Mineirao stadium downloaded from the Google SketchUp 3D Warehouse library. *(c)* Closer view on one of the 88 PV generators. *(d)* Heat map of the shading affecting some of the PV generators.

The solar installation consists of 88 PV generators composed of 14x4 PV panels. Each PV generator presents a different orientation and different shading surroundings. Consequently, the shading losses needed to be evaluated for each of these 88 sections.

The 3D model of the stadium was downloaded from Google Earth. The detailed shading assessment for each PV generator allowed to optimize the PV panels connection architecture and the inverter set up in order to minimize the energy output losses [20].

3.4 PV plant with 2-axes trackers on uneven ground

Figure 9 shows a PV plant that was projected in Spain (close to Madrid) with 18 2-axes trackers on an uneven ground. A detailed 3D mesh for the ground was obtained from a topographical study whose results where first treated with AutoCAD, and then imported into 3DPV. There is an electrical pylon on the field, whose shading is complex to evaluate, and whose 3D model was imported from Google SketchUp 3D Warehouse library. The shadows cast by the electrical pylon on each one of the PV trackers were evaluated with a high spatial resolution.

This study has allowed optimizing the PV plant layout. To our knowledge, no commercial software would allow to do so in such a situation.

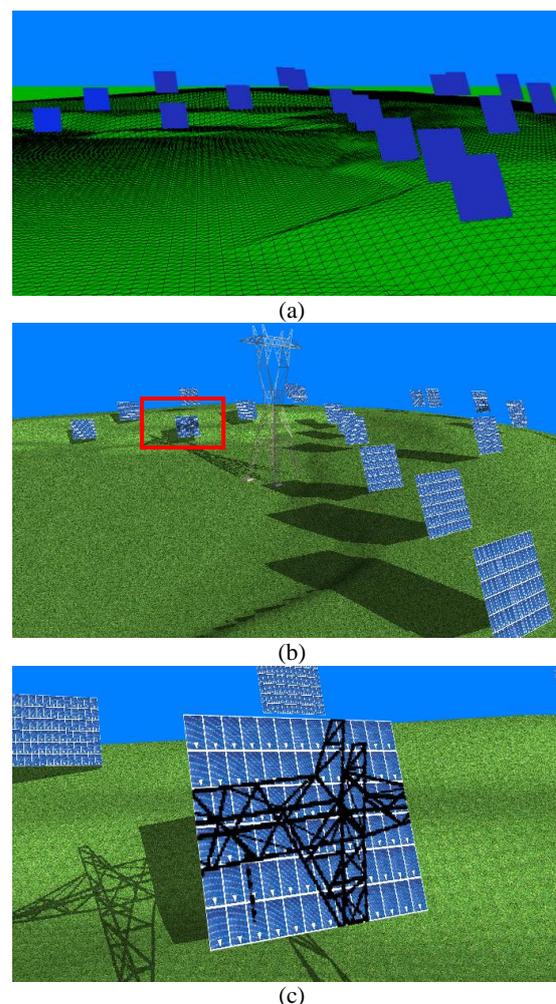

**Figure 9:** PV plant with 2-axes trackers on an uneven ground. (a) The 3D mesh for the ground was obtained from a topographical study whose results where first treated with AutoCAD, and then imported into 3DPV. (b) View of the PV generators and the electrical pylon. (c) Closer view of the shadows cast by the electrical pylon on one particular PV generator.

## 4  DISCUSSION

The time required to evaluate complex shading situations is greatly reduced when using the GPU. The shading evaluation corresponding to figure 5 is the most complex one, in terms of the number of elements contained in the 3D model (80,000 triangles). For this case, we carried out a shading evaluation for a period of one year, with a temporal resolution of 1 simulation each 10 minutes. The whole shading scene was then evaluated for around 26,000 time steps (one year of daylight moments at 10-min temporal resolution). The spatial resolution was set to one area per PV cell. This represents some 1,400 for the whole PV generator. This represents some 80,000 x 26,000 x 1,400 or around 3E12 shading evaluations. The computation was completed after a time of 8 minutes using a small laptop computer.

This tool is implemented using *WebGL*, and it is directly run from any Web browser supporting *HTML5*, such as *Google Chrome* [21] or *Mozilla Firefox* [22]. Therefore, it does not need to be installed locally on a client computer. Its full online integration online also means that it can make use of 3D objects libraries available online such as *Google Earth* and *Google SketchUp 3D Warehouse*.

This tool has initially been implemented for the PV sector, but it can also be used for other kinds of applications, such as the evaluation of shading affecting solar thermal panels, or buildings in general.

In its present state, our shading simulations only consider the effect of shading on direct solar irradiation. In reality, shading can also obstruct one part of the diffuse component of solar irradiation. In most cases, this influence is nevertheless of very little energetic relevance. Although, it would be technically possible to take them into account, using a complementary method called shadow volume [23].

## 5  CONCLUSION

We have developed a tool, *3DPV*, which is capable of evaluating the PV energy output losses caused by complex shading environments.

We have applied it to the evaluation of shading scenes in BIPV and PV plants.

This tool is implemented using *WebGL*, and it is directly run from any Web browser supporting *HTML5*. Therefore, it does not need to be installed locally on a client computer, and it can be directly used online. Its full integration online also means that it can make use of 3D objects libraries available online.

## ACKNOWLEDGMENTS

We are grateful to Catherine Praile, who carried out a great proofreading job, as usual. This work has been partially supported by the European Commission within the project PV CROPS [24] (Photovoltaic Cost r€duction, Reliability, Operational performance, Prediction and Simulation) under the 7th Framework Program (Grant Agreement nº 308468).